\documentclass{article}
\usepackage{spconf,amsmath,graphicx}   
\usepackage{adjustbox}   

\usepackage{enumitem}   
\usepackage[colorlinks]{hyperref}   
\setlist{nosep, leftmargin=14pt}

\usepackage{mwe} 

\title{A Wavelet Guided Attention Module for Skin Cancer Classification with Gradient-based Feature Fusion}
\name{ {Ayush Roy$^{1,\dagger}$ \qquad \newline \hspace{0.1 cm}   
      Sujan Sarkar$^{2,\dagger}$ \qquad \newline \hspace{0.1 cm}   
      Sohom Ghosal$^{3,\dagger}$ } \thanks{$^{\dagger}$ These authors contributed equally to this work.} \qquad \newline \hspace{0.1 cm} 
      Dmitrii Kaplun$^{4,5,\star}$ \thanks{$^{\star}$ Corresponding author.} \qquad   \newline   \hspace{0.1 cm}  
      Asya Lyanova$^{5}$ \qquad \newline \hspace{0.1 cm}   
      Ram Sarkar$^{2}$}   

\address{$^1$ Department of Electrical Engineering, Jadavpur University, Kolkata, India. \\
        $^2$ Department of Computer Science and Engineering, Jadavpur University, Kolkata, India. \\
        $^3$ Department of Electronics and Communication Engineering, International Institute of \\ Information Technology, Naya Raipur, Chhattisgarh, India. \\     
       $^4$ Artificial Intelligence Research Institute, China University of Mining and Technology,  Xuzhou, China.\\
       $^5$ Department of Automation and Control Processes, Saint Petersburg Electrotechnical \\
        University "LETI", Saint Petersburg, Russia.  }
\begin{document}
\maketitle
\begin{abstract}
Skin cancer is a highly dangerous type of cancer that requires an accurate diagnosis from experienced physicians. To help physicians diagnose skin cancer more efficiently, a computer-aided diagnosis (CAD) system can be very helpful. In this paper, we propose a novel model, which uses a novel attention mechanism to pinpoint the differences in features across the spatial dimensions and symmetry of the lesion, thereby focusing on the dissimilarities of various classes based on symmetry, uniformity in texture and color, etc. Additionally, to take into account the variations in the boundaries of the lesions for different classes, we employ a gradient-based fusion of wavelet and soft attention-aided features to extract boundary information of skin lesions. We have tested our model on the multi-class and highly class-imbalanced dataset, called HAM10000, and achieved promising results, with a 91.17\% F1-score and 90.75\% accuracy. The code is made available at: \url{https://github.com/AyushRoy2001/WAGF-Fusion}.        
\end{abstract}

\section{Introduction}
\par With thousands of deaths annually, skin cancer has become one of the most deadly cancers globally in recent times. Sun exposure is the primary cause of skin cancer. Skin cancer can be classified into several types, with melanoma and non-melanoma being the major ones \cite{elgamal2013automatic}. Melanoma has a very low survival rate, making early diagnosis crucial to prevent it from spreading throughout the body and causing painful death \cite{khan2019classification}. There are two main diagnostic methods for skin cancer: dermoscopy and skin biopsy. Dermoscopy requires a well-trained physician with experience and visual ability, while skin biopsy involves taking a sample of skin from the patient's body, which can be a slow and painful process. These difficulties have spurred researchers in the field of artificial intelligence (AI) to create Computer-Aided Diagnosis (CAD) systems capable of precise skin cancer classification. However, due to the inter-class similarity and intra-class dissimilarity among different types of skin cancer, classifying skin cancer using medical image processing is a challenging issue. 

\textbf{Related Work:} Yao et al. \cite{yao2021single} proposed a deep convolutional neural network (CNN) with a new loss function, called Multi-Weighted New Loss (MWNL) function to tackle the issues with small and imbalanced datasets. Most of the skin image datasets are highly imbalanced. Different data augmentation techniques have been introduced to overcome this problem. There is a data augmentation technique with high performance but low cost which is developed for skin cancer screening devices that can be distributed in non-developed communities \cite{shen2022low}. Raw skin image datasets have images with many artifacts and low contrast. Some preprocessing techniques like, noise removal, contrast enhancement technique, and resizing have been proposed to assist the classifier to perform efficiently \cite{charan2020method}. Mridha et al. \cite{mridha2023interpretable} used two optimization functions (Adam, RMSprop) and three activation functions (Swish, ReLU, Tanh) with a CNN model in the training phase so that the classifier performed with higher accuracy. Iqbal et al. \cite{iqbal2021automated} designed a deep CNN consisting of a total of 63 convolution layers to deal with the multi-class classification of skin cancer, being aware of inter-class similarities and intra-class variances, but their model was insufficiently capable of addressing these issues. To deal with inter-class similarities and intra-class variances, A loss function, called Margin-Aware Adaptive-Weighted (MAAW) loss, has been incorporated with the training model by Roy et al. in \cite{roy2023margin}. However, it does not provide satisfactory performance. As a result, it is evident from this literature review that there is a pressing need for improvement of the skin cancer classification models.

\textbf{Contributions:} According to \cite{astierskincancer}, melanoma can be differentiated from a normal mole or other classes by observing its diameter, uniformity in color, shape, symmetry, etc.  In light of this fact, the current research suggests a unique architecture that, by extracting semantic differences in characteristics across spatial dimensions and obtaining boundary and symmetry information, can reliably categorize skin cancer images. Since it might be challenging to pinpoint the precise location of skin lesions, the boundary information of lesions has been strengthened in this research by a gradient-based fusion of the wavelet and attention-aided features, leading to an efficient classification model.  

\section{Methodology}
\begin{figure}
    \centering
    \includegraphics[width=0.95\linewidth]{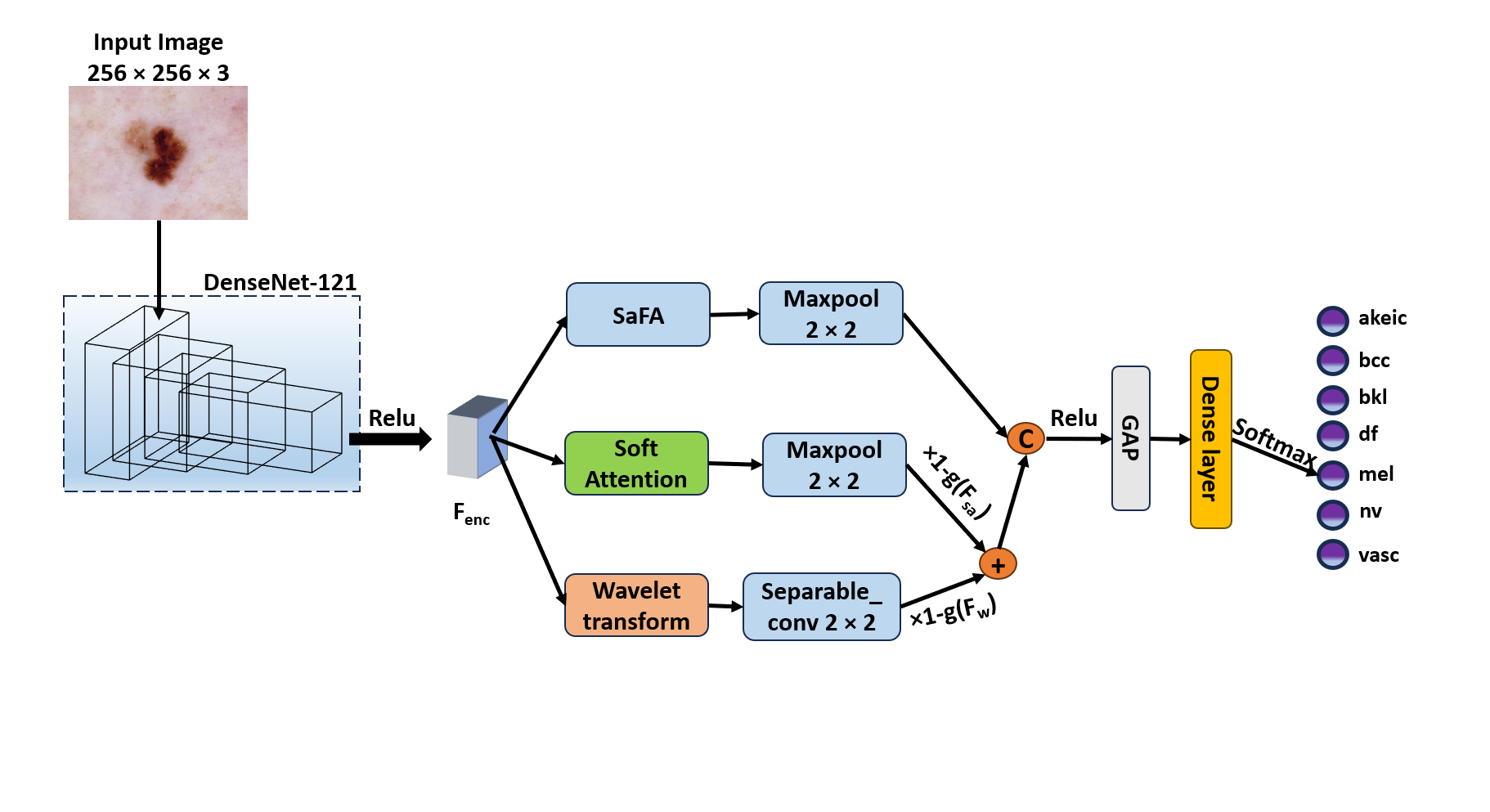}    
    \caption{The proposed classification model utilizes the DenseNet-121 as the backbone for feature extraction. The extracted feature $F_{enc}$ is treated with Soft Attention and the SaFA module. A gradient-based fusion of the wavelet and attention-aided features reinforces the boundary information of the lesions.} 
    \label{fig:arch_proposed}   
\end{figure}

To create a lesion-aware skin cancer type classification model, we have introduced the Symmetry-aware Feature Attention (SaFA) module and a gradient-based feature fusion mechanism to combine the wavelet and convolution features in addition to the existing soft attention. We have used DenseNet-121 as the backbone for feature extraction. The extracted feature, $F_{enc}$, is processed with the attention modules and then flattened using a Global Average Pooling layer (GAP). This flattened feature is then passed on to a dense layer with softmax activation to predict the classes. A block diagram overview of the proposed model can be seen in Fig. \ref{fig:arch_proposed}. 

\subsection{Gradient-based Feature Fusion} 
Wavelet transformation is a technique that is used for image analysis and compression. It detects the edges in an image by considering high frequency components, after transforming the image to its frequency distribution. In our proposed model, we use one of the most effective wavelet transformations, the Haar wavelet transformation, which extracts wavelet features denoted as $F_{w}$ from $F_{enc}$. These features include the boundary information of the lesion, which provides necessary insights to the model for differentiating different classes. To improve the feature extraction process, we dynamically apply an adaptive gradient-based feature fusion of $F_{w}$ and the Soft Attention feature $F_{sa}$. The fusion process uses the normalized backpropagation gradients of $F_{w}$, denoted as $g(F_{w})$, and that of $F_{sa}$, denoted as $g(F_{sa})$, as weights for the fusion process. Equation \ref{eq:fusion} shows how this works. A higher gradient means that the feature negatively impacts the model's performance, so it is given less weightage in $F_{fuse}$ and vice versa. With this dynamic weight allocation, feature fusion is more optimized and representable without increasing the parameters by introducing learnable weights.

\begin{equation}
    F_{fuse} = (1-g(F_{w})) \times F_{w} + (1-g(F_{sa})) \times F_{sa}
    \label{eq:fusion}
\end{equation}

This approach ensures that the boundary information is added to the encoded feature before passing it on to the classification layer.

\subsection{Symmetry-aware Feature Attention Module}
\begin{figure}
    \centering
    \includegraphics[width=0.95\linewidth]{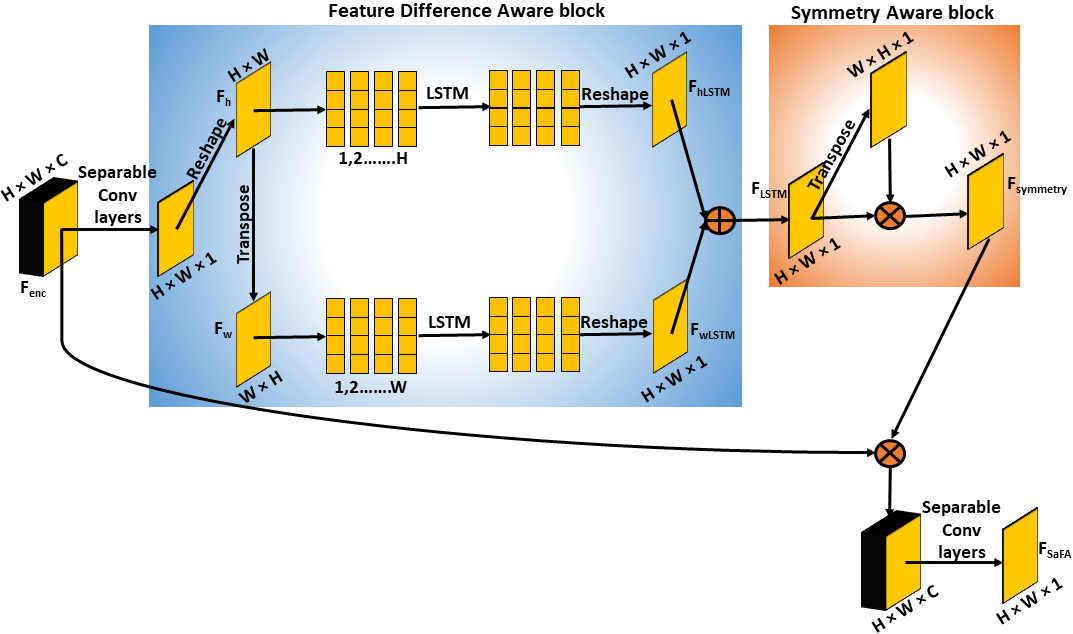}
    \caption{Symmetry-aware Feature Attention (SaFA) module.} 
    \label{fig:attention}
\end{figure}

The main purpose of the SaFA module is to extract information related to the symmetry of the lesions and the changes in features across the spatial dimensions of the regions, i.e., the semantic changes. The SaFA module consists of two main blocks, the Symmetry-aware Block (SaB) and the Feature Difference-aware Block (FDaB). 

The FDaB inputs the $F_{enc}$ of dimensions $H \times W \times C$ and converts it to a dimension of $H \times W \times 1$ using three separable convolution layers of filters 256, 64, and 1 consecutively. This feature map is then reshaped to $F_{h}$ and $F_{w}$ of dimensions $H \times W$ and $W \times H$ respectively. Since LSTM (Long Short-Term Memory) based network is capable of understanding and preserving temporal dependencies, we have used them for capturing the changes in semantic features across a particular dimension. Here, $F_{h}$ and $F_{w}$ serve as input to the LSTM layer, where $H$ and $W$ act as the timestamps respectively, as shown in Fig. \ref{fig:attention}. Especially, the main purpose of the LSTM layers is to identify the change in features as we move through the spatial dimension along $H$ and $W$ of the feature maps $F_{h}$ and $F_{w}$, respectively. The outputs of both these LSTM layers are reshaped to dimension $H \times W \times 1$ for producing $F_{hLSTM}$ and $F_{wLSTM}$. These two feature maps are added together to generate $F_{LSTM}$.

The SaB takes $F_{LSTM}$ as input to produce $F_{symmetry}$ as shown in Eq. \ref{eq:symmetry}, where $\otimes$ represents pixel-wise feature similarity. This symmetry-aware feature, $F_{symmetry}$, is treated with separable convolution layers similar to the FDaB followed by a sigmoid activation function to produce $F_{attn}$.

\begin{equation}
    F_{symmetry} = F_{LSTM} \otimes F_{LSTM}^T
    \label{eq:symmetry}
\end{equation}

\section{Results}

\subsection{Experimental Setup}
Skin lesions can be classified into different types, such as Actinic keratoses and intraepithelial carcinoma/Bowen's disease (akiec), basal cell carcinoma (bcc), benign keratosis-like lesions (bkl), dermatofibroma (df), melanoma (mel), melanocytic nevi (nv), and vascular lesions (vasc). The main differences between these types are their symmetry, uniformity of features (such as color and texture), and boundary region. For our research, we have trained and evaluated our model using the HAM10000 dataset. This dataset contains 10,015 dermatoscopic images of both common and rare skin lesions, divided into seven classes. The dataset is highly class-imbalanced, which presents challenges to achieving state-of-the-art (SOTA) results. For our experiments, we have used 1103 images out of 10015 original images for testing. We have augmented rest 8912 original images to 52000 images and used 70\% of augmented images for training and 30\% for validation, all of dimensions $256 \times 256 \times 3$. We have used a learning rate of 0.01, the Adam optimizer, and a batch size of 64 for all of our experiments. We have applied the cross-entropy loss function for training the model over 25 epochs and evaluated it using standard metrics in TensorFlow, including accuracy, precision, recall, and F1-score on an NVIDIA TESLA P100 GPU.

\subsection{Ablation study}
For selecting the backbone, we have trained three models, MobileNet-v2, DenseNet-121, and ResNet50. Among these, DenseNet-121 gives the best results as shown in Table \ref{ablation}. Using DenseNet-121 as the backbone, we have performed more experiments to figure out the best architectural configuration. These experiments are:

(i) DenseNet-121 + Soft Attention

(ii) DenseNet-121 + Soft Attention + Gradient-based feature fusion

(iii) The proposed model ((ii) + SaFA module)

\begin{figure}
    \centering
    \includegraphics[width=\linewidth]{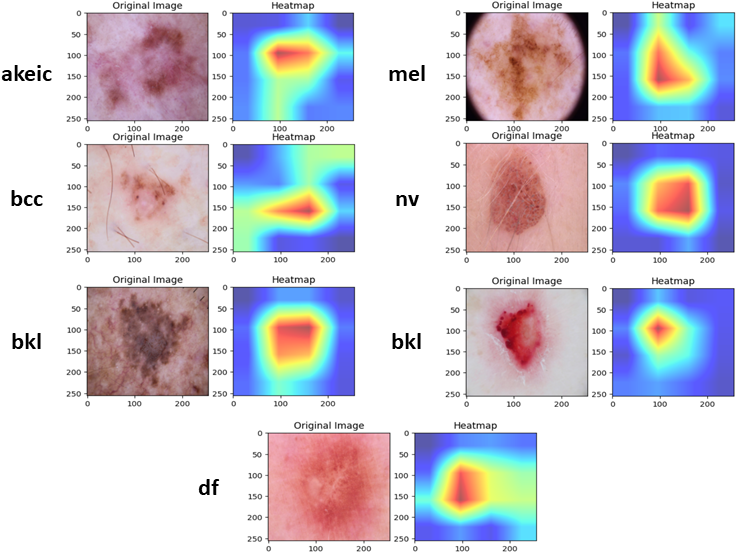}
    \caption{Heatmap of the SaFA module for all seven classes.} 
    \label{fig:heat}
\end{figure}

In Table \ref{ablation}, it is seen that the introduction of gradient-based feature fusion provides a necessary boost to performance by incorporating the boundary features of the lesions. Furthermore, the SaFA module further enhances the performance by spotting the symmetry and the semantic feature changes of the lesions as shown in Fig. \ref{fig:heat}, where the attention map of the SaFA module is shown.

\begin{table}[h!]
    \centering
    \caption{Ablation study on the HAM10000 dataset \cite{tschandl2018ham10000} for selecting the best model configuration.}    
    \begin{adjustbox}{max width=\linewidth}
    \begin{tabular}{lcccc}
        \hline
        \textbf{Model} & \textbf{Accuracy (\%)} & \textbf{Precision (\%)} & \textbf{Recall (\%)} & \textbf{F1-score (\%)}\\      
        \hline
        MobileNet-v2 & 79.03 & 78.78 & 79.39 & 78.60\\ 
        ResNet50 & 84.66 & 84.58 & 85.72 & 83.31\\ 
        DenseNet-121 & 85.89 & 85.22 & 85.87 & 84.85\\ 
        (i) & 87.21 & 87.71 & 86.76 & 87.43\\
        (ii) & 89.48 & 90.05 & 89.78 & 90.08\\
        \textbf{(iii)} & \textbf{90.75} & \textbf{90.83} & \textbf{90.75} & \textbf{91.17}\\
        \hline
    \label{ablation}
    \end{tabular}
    \end{adjustbox}
\end{table}   

\subsection{Comparison with SOTA}   

Our proposed model outperforms existing methods in terms of accuracy and F1-score, as shown in Table \ref{tab:SOTA}.   

\begin{table}[h!]      
\centering
 \caption{Performance comparison between proposed model and existing methods on the HAM10000 dataset \cite{tschandl2018ham10000}.}   

 \begin{adjustbox}{max width=\linewidth}
 \begin{tabular}{lccccc}   
 \hline
 \textbf{Model} & \textbf{Accuracy (\%)} & \textbf{Precision (\%)} & \textbf{Recall (\%)} & \textbf{F1-score (\%)} \\  
 \hline  
 Yao et al. \cite{yao2021single} & 86.40 & - & - & - \\
 Iqbal et al. \cite{iqbal2021automated} & 88.80 & 88.80 & 89.10 & 90.50  \\        
 Shen et al. \cite{shen2022low} & 85.30 & - & - & - \\    
 Sai Charan et al. \cite{charan2020method} & 88.60 & - & - & - \\
 Mridha et al. \cite{mridha2023interpretable} & 82.00 & - & - & -  \\   
 Roy et al. \cite{roy2023margin} & - & 70.60 & 70.20 & 72.30 \\   
 \textbf{Proposed} & \textbf{90.75} & \textbf{90.83} & \textbf{90.75} &  \textbf{91.17}\\ 
 \hline
 \end{tabular}
 \end{adjustbox}
 \label{tab:SOTA}   
\end{table}  

\begin{figure}
    \centering
    \includegraphics[width=0.75\linewidth]{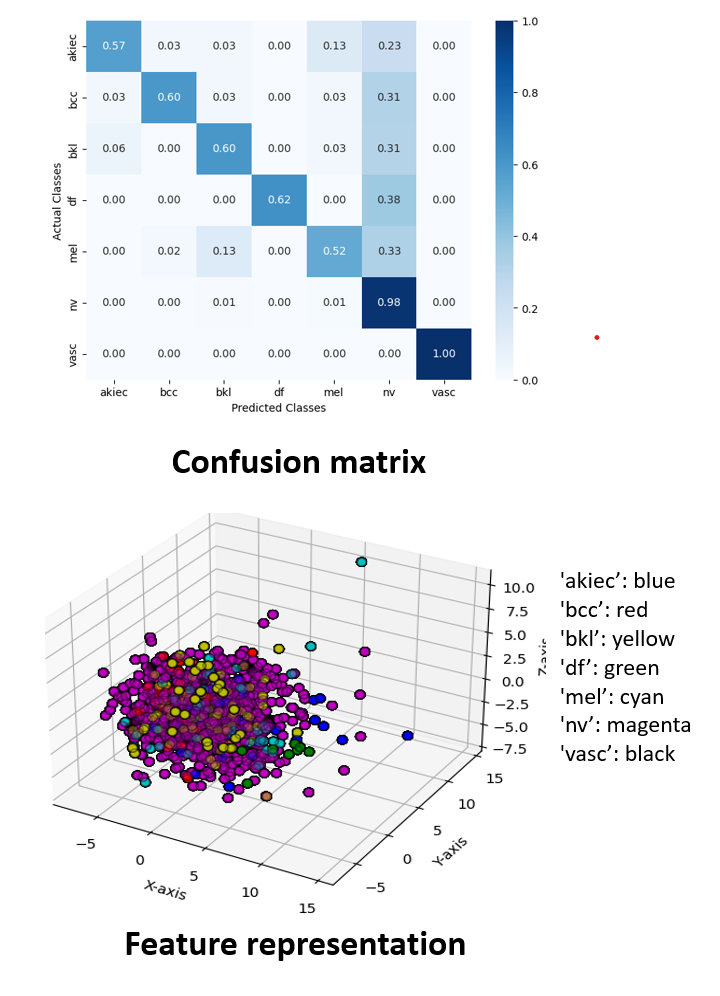}    
    \caption{Confusion matrix and feature representation of the proposed model.} 
    \label{fig:results}
\end{figure}

In Fig. \ref{fig:results}, we can see that all seven skin diseases have been efficiently classified, by observing the confusion matrix given by our proposed model on HAM10000. Also, the distribution of the flattened features of various classes is shown in a 3D scatter plot to visualize the differentiability of the classes.

\section{Conclusion}
We have developed a novel architecture to classify various types of skin cancer by analyzing skin images with lesions. The model extracts meaningful features from the images with a focus on the variations in the lesion boundaries by incorporating the boundary information from wavelet features. Additionally, an attention mechanism is used to account for the variations of texture and color across the region of interest, i.e., the changes in semantic features of the lesions and the symmetry of the lesion. The model uses a gradient-based fusion of the wavelet features with the attention-aided feature by assigning an appropriate weightage to the more relevant information. Despite the results achieved by the proposed model, the recall rates of skin cancer types like melanoma and other minority classes need to be improved for real-world usability.   
We plan to improve the recalls of melanoma and minority classes explore the model's performance across other medical image modalities and streamline it to create a lightweight model.

\bibliographystyle{IEEEbib}
\bibliography{strings,refs}

\end{document}